\documentstyle[twocolumn,aps,pre,epsfig]{revtex}
\begin {document}
\wideabs{
\title{Primary Sequences of Protein-Like Copolymers:
Levy Flight Type Long Range Correlations}
\author{Elena N. Govorun,$^{*1}$ Victor A. Ivanov,$^{1}$ Alexei R. Khokhlov,$^{1}$
Pavel G. Khalatur,$^{2}$ Alexander L. Borovinsky,$^{3}$ Alexander
Yu. Grosberg,$^{3,4}$}
\address{$^{1}$Physics Department, Moscow State University, Moscow, 119899 Russia \\
$^{2}$Department of Physical Chemistry, Tver State University,
Tver 170002, Russia \\
$^{3}$Department of Physics, University of Minnesota, Minneapolis,
MN 55455, USA \\ $^{4}$ Institute of Biochemical Physics, Russian
Academy of Sciences, Moscow 117977, Russia }
\date{\today}

\maketitle
\begin{abstract}
We consider the statistical properties of primary sequences of
two-letter $HP$ copolymers ($H$ for hydrophobic and $P$ for polar)
designed to have water soluble globular conformations with $H$
monomers shielded from water inside the shell of $P$ monomers. We
show, both by computer simulations and by exact analytical
calculation, that for large globules and flexible polymers such
sequences exhibit long-range correlations which can be described
by Levy-flight statistics.

\

\end{abstract}

}

In refs \cite{PhysA,PhRev}, a new approach to the design of
specific primary sequences for the $HP$-copolymers consisting of
monomeric units of two types (hydrophobic $H$ and polar $P$) has
been proposed by some of the present authors. Unlike some other
methods of sequence design known in the literature (see review
\cite{RMF} and references therein), the approach in question does
not aim to mimic folding into one particular conformation. The
goal is to model simpler and more robust property of proteins,
such as their ability to stay dissolved and shield their
hydrophobic monomers from water. The essence of this approach is
illustrated in Fig.\ 1. We start with an arbitrary
computer-generated globular conformation of a homopolymer chain
(formed due to the strong attraction of monomer units, Fig.\ 1a)
and perform a "coloring" procedure: monomer units in the core of
the globule (having many neighbors) are set to be $H$-units while
monomer units belonging to a globular surface (where the number of
neighbors is smaller) are assigned to be of $P$-type, Fig.\ 1b.
Then the obtained primary sequence is fixed, uniform attraction of
monomer units is removed and newly generated $HP$-copolymer is
ready for the further investigation (Fig.\ 1c). Thus obtained
macromolecules are {\em protein-like} in the sense that they mimic
segregation of globule into hydrophobic core and stabilizing
hydrophilic envelope.  The properties of protein-like copolymers
were examined in \cite{PhysA,PhRev,ABpl,Symp}; see also
\cite{EXPERIMENT,experim2} on possible ways of experimental
realization.

In this Letter, we address correlations between $H$- and $P$-units
along the protein-like sequences.  This may shed light on the
conditions which must be met by the sequence to provide for the
water solubility of globules - the issue of great potential
relevance to our understanding of early evolution. We show, both
by computer simulations and by exact analytical calculation, that
correlations have a long-range character. More specifically, for
the simple model of flexible polymer, they belong to the so-called
Levy flight statistics.

To begin with, statistical properties of protein-like
$HP$-sequences can be assessed computationally by the method
similar to that used by Stanley et al \cite{St1,St2} in their
search for long-range correlations in DNA sequences.  We choose
the "window" of length $\ell$, move it step by step along the
generated $HP$-sequence, and at each step count the number of $H$
units inside the window.  This number, which we write as
$\sum_{i=j}^{j+\ell} u_{i}$ is a random variable, depending on the
position $j$ of the window along the sequence; here $u_i$ is the
variable associated with every monomer $i$ such, that $u_i=1$ if
monomer $i$ is $H$ and $u_i=0$ if it is $P$.  This random variable
has certain distribution. Its average is determined by the overall
sequence composition (total numbers of $H$-
and $P$- monomers), and its dispersion is easy to calculate:
\begin{equation}
D_{\ell}^2 = \sum_{i,j=k}^{k+\ell} \left( \left< u_i u_j \right> -
\left< u_i \right> \left< u_j \right> \right) \ . \label{eq:dispersion}
\end{equation}
For a completely random $HP$-sequence, the value of $D_{\ell}$ scales as
${\ell}^{1/2}$ with the window width ${\ell}$. The dependence $D_{\ell}
\sim {\ell}^{\alpha}$ with $\alpha > 1/2$ would then manifest the
existence of long-range correlations.

\begin{figure}
\epsfig{file=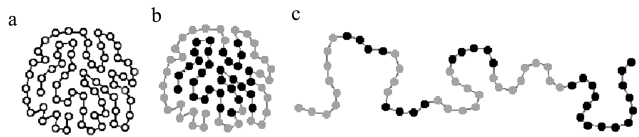, width=8.6cm} \caption{Sequence design scheme
for protein-like copolymers: (a) homopolymer globule; (b) the same
globule after the coloring procedure; (c) protein-like copolymer
in the coil state.}
\end{figure}

The result of such calculation for averaging over 2000 independent
protein-like $HP$-sequences of $N=1024$ monomer units with $1:1$
composition (obtained as in ref.\ \cite{PhRev}) is presented in
Fig.\ 2, squares. For comparison, in the same figure 2 the
data for two other types of sequences (averaged over 2000
independent species) are shown. One of them is purely random $1:1$
sequence, it demonstrates $D_{\ell} \sim {\ell}^{1/2}$ scaling.
Comparing this curve with Monte Carlo results we see immediately that
protein-like sequence is not random and some correlations do exist
in it. Thus, it is interesting to compare the squares in Fig.\ 2 and
the dashed curve showing data for the sequence which was called "random-block"
in refs \cite{PhysA,PhRev,ABpl,Symp}: the lengths of $H$ and
$P$-blocks in a sequence are determined by the Poisson
distributions adjusted to achieve the same $1:1$ composition and
the same "degree of blockiness" (average block length) as for
protein-like $HP$-copolymer. This sequence exhibits a somewhat
more rapid variation of $D_{\ell}$ at small ${\ell}$, but
ultimately the law $D_{\ell} \sim {\ell}^{1/2}$ is obeyed for
large values of ${\ell}$. Nevertheless, this random-block model is
also seen to be unsatisfactory for the statistical behavior of
protein-like sequence throughout the interval of ${\ell}$
examined, $2 < {\ell} < 500$.  Although the data do not fit
accurately to any power law $D_{\ell} \sim {\ell}^{\alpha}$, the
slope of the observed $D_{\ell}$ dependence corresponds to
$\alpha$ significantly larger than $1/2$, up to about $0.85$, thus
indicating pronounced long-range correlations in protein-like
sequence.  In what follows we present analytical theory which
produces curve $D_{\ell}$ in Fig.\ 2 in complete agreement with
observations.

First of all, let us turn to the origin of long-range correlations
in the primary sequences for $HP$-copolymers generated via the
procedure illustrated in Fig.\ 1. Conceptually, this problem is
fairly easy to address: since sequence in this scheme is uniquely
determined by the parent conformation, the statistics of sequences
reflects nothing but the statistics of parent conformations,
which, in turn, is well understood. Indeed, the coloring procedure
(Fig.\ 1b) operates in dense globular conformation.  Since we
consider very compact parent conformations, the statistics of
polymer chain conformations inside the globule is ideal (Gaussian)
according to the well-known Flory theorem \cite{uchebnik}.
Therefore, all the statistical properties of parental
conformations, including the correlations in the primary sequences
produced by coloring procedure, can be derived via the solution of
diffusion equation for random walks with appropriate boundary
conditions.

To understand the fractal aspect of the sequences, it is
convenient to concentrate on their uninterrupted homo-colored
sections and on the points of connection between them. Our
coloring procedure (Fig.\ 1) introduces a separation sphere of
radius $R^{\ast} < R$, such that all the units which in the parent
conformation are confined inside this sphere are of $H$-type and

the units belonging to the shell layer $R^{\ast} < r < R$ are of
$P$-type. Therefore, homo-colored section is produced in our model
by the chain section of the parent conformation placed entirely in
either internal or shell regions of the globule. The probability
to have an uninterrupted succession of some $k$ of $H$-monomer
units in the sequence is equal to the probability that Gaussian
(due to Flory theorem) polymer has a loop of $k$ monomer units
entirely confined in $H$-region with ends on the separation
surface. Similarly, the probability to have an uninterrupted
succession of some $k$ of $P$-monomer units in the sequence is
equal to the probability that ideal parental conformation has a
loop of $k$ monomer units confined within the shell $P$-region,
again with ends on the separation surface.

\begin{figure}
\centerline{\epsfig{file=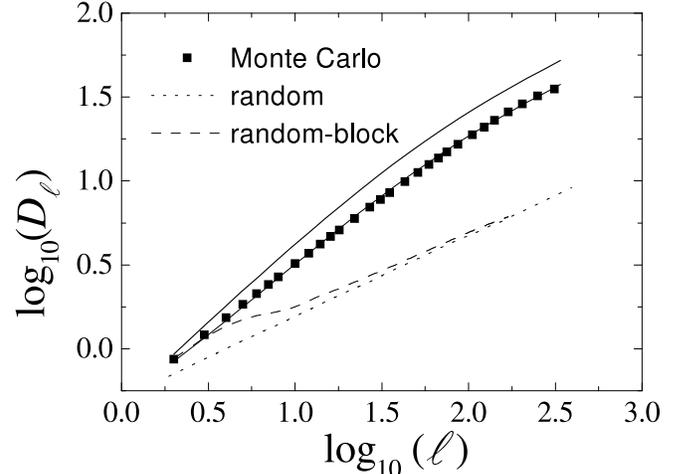, width=8.6cm}}
\caption{Dispersion of the number of $H$-units in the fragment of
sequence of size $\ell$ for protein-like $HP$-sequence, random
copolymer and random-block copolymer. Results of analytical theory
for protein-like sequence are shown both for continuous approximation
by thick solid line and for discrete approximation by thin solid line
(see explanation around eq.\ (\ref{eq:finalD})). Corresponding Monte
Carlo results are presented by squares. There is no adjustable parameters
involved in the fit, length scale $a$ is uniquely determined by
the geometry of bond fluctuation model \protect\cite{expl}}
\end{figure}

To address probability distributions $P_H(k)$ and $P_P(k)$, we
begin with simple physical arguments yielding
\begin{equation}
P_{H,P}(k) \simeq \cases{  k^{-3/2}, & $1< k < \left(
\frac{d_{H,P}}{a} \right)^2$; \cr \left( \frac{a}{d_{H,P}}
\right)^{3} e^{- \lambda_{H,P} k}, & $k > \left( \frac{d_{H,P}}{a}
\right)^2$. \cr  }\label{eq:qualitative}
\end{equation}
The upper asymptotic form is valid for short polymer loops, when
neither curvature of the separating boundary nor overall globule
shape play any role.  In this regime, $P(k)$ is simply the
probability for a random walk to start at the planar wall and to
return to it for the first time after $k$ steps.  This is
classical probabilistic "first return" problem, for which the
$\sim k^{-3/2}$ answer is well known \cite{Feller}. This scaling
is valid for loop sizes $a \sqrt{ k }$ much larger than unity but
smaller than the relevant characteristic length scale,
$d_H=R^{\ast}$ for the $H$-loops inside the inner sphere, or
$d_P=R-R^{\ast}$ for the $P$-loops in the spherical shell.  The
second asymptotic form in equation (\ref{eq:qualitative})
indicates that for long polymer loops the function $P(k)$ decays
exponentially. It is easier to explain this in terms of polymer
statistics: to confine a polymer chain of $k$ monomer units in a
cavity costs some entropy $\Delta S$, at $a \sqrt{k} \gg d$ this
entropy goes linearly with $k$, making the probability,
$\exp(\Delta S)$, exponential in $k$.

Let us now look closer at the cross-over values of $k$.  In order
to achieve the $1:1$ composition, $R^{\ast}$ must be chosen such
that volumes of internal $H$- and shell $P$-regions are the same,
which means
$R^{\ast} = 2^{-1/3} R \approx 0.8 R$. The volume fraction of
polymer units in a globule, $\phi$, is controlled by the energy of
interactions of monomer units used to prepare parental
conformation. It is clear that 
$R \approx 0.6 a N^{1/3} \phi^{-1}$. Therefore, $H$-loops remain
in the power law long range correlation regime up to the length $k
< 0.24 N^{2/3} \phi^{-2}$, while for $P$-loops this cross-over
occurs somewhat earlier: $k < 0.015 N^{2/3} \phi^{-2}$. Thus, we
predict that there should be over a decade of length scales in
which $H$-loops are still long range correlated, while only short
range correlations remain in the $P$-loops.

The result (\ref{eq:qualitative}) is sufficient to explain
qualitatively correlations in protein-like sequences, including
the data shown in Fig.\ 2.  Indeed, according to our discussion,
protein-like sequence can be thought of as an alternating
succession of $H$- and $P$-stretches, with lengths of stretches
taken independently from the corresponding distributions $P_H(k)$
and $P_P(k)$. This mathematical scheme is called a Levy flight
\cite{Levy}. We conclude that the long-range correlations in the
primary sequences of protein-like copolymers are described by Levy
flight statistics.  Furthermore, for the $k^{-3/2}$ behavior of
$P(k)$, the averaged block length diverges, and, therefore, the
value of $D_{\ell}$ in the power law regime is controlled by the
longest block, yielding $D_{\ell} \sim \ell$, or $\alpha =1$. This
is true as long as both $H$- and $P$- loops remain in fractal
regime. On the other hand, when all loops cross-over to
exponential distribution, $D_{\ell}$ crosses over to $\alpha
=1/2$:
\begin{equation}
D_{\ell} \simeq \cases{ \ell \ , & for  $1<  \ell < 0.015 N^{2/3}
\phi^{-2}$ \cr \ell^{1/2} \ , & for $\ell > 0.24 N^{2/3}
\phi^{-2}$ \cr } \label{eq:qualitativeD}
\end{equation}
The cross-over region for $D_{\ell}$ is very broad, it corresponds
to the situation in which $P$ loops are already "large, while
$H$-loops are still "small."  Both $\alpha=1$ and $\alpha =1/2$
limits and wide cross-over agree qualitatively well with
computational data, Fig.\ 2. This motivates more careful theory,
in which instead of scaling estimates (\ref{eq:qualitative}) and
(\ref{eq:qualitativeD}), the expressions in terms of infinite
series, suitable for numerical calculation, is obtained.

To develop full analytical theory, it is convenient to use the
random walk terminology to describe parent conformation. In this
language, for instance, $P_H(k)$ is the probability that the
random walker enters a sphere of the radius $R^{\ast}$ and then
arrives back to the boundary {\em for the first time} after "time"
$k$.  Recall that the statistical weight of all random walk
trajectories starting at the point $\vec{r}_0$ and arriving after
$k$ steps at the point $\vec{r}$, $G\left( \vec{r},k \left|
\vec{r}_0 \right. \right)$, obeys the diffusion equation
\begin{equation}
\frac{ \partial G\left( \vec{r},k \left| \vec{r}_0 \right. \right)
}{\partial k } = \frac{a^2}{6} \Delta G\left( \vec{r},k \left| \vec{r}_0
\right. \right) + \delta (k) \delta (\vec{r} - \vec{r}_0 )
\label{eq:Diffusion_equation} \end{equation}
where $a^2$ is the mean square length of one step (the squared size of
one monomer unit along the chain). To introduce the condition of first
return we have to say that the walker never touches the boundary, which
is achieved by imposing the boundary condition
\begin{equation}
\left. G\left( \vec{r},k \left| \vec{r}_0 \right. \right) \right|
_{\left| \vec{r} \right| = R^{\ast}} = 0 \ . \label{eq:boundary1}
\end{equation}
The probability distribution of the "first return times" in terms of $G$
is then given as the time-dependent flux of diffusing particles through
the absorbing wall:
\begin{equation}
P_H(k) = \left| \oint d \sigma \left. \frac{a^2}{6} \frac{\partial G
}{\partial r} \right|_{r = R^{\ast}} \right| \ , \label{eq:flux}
\end{equation}
where $\partial / \partial r$ means the component of gradient normal to
the surface, integration is performed over the closed separating
surface, and the absolute value is written to avoid thinking about the
direction of the flux.  The normalization condition $\int P_H(k) dk =1$
is guaranteed by the fact that all diffusing particles eventually leave
through the surface.  As regards $\vec{r}_0$, it should be taken within
a distance of order $a$ from the separating $R^{\ast}$ surface. The
problem thus formulated, including equations
(\ref{eq:Diffusion_equation}-\ref{eq:flux}), is easy to solve:  we write
$G$ in terms of bilinear expansion $G = \sum_n e^{k \lambda_n} \psi_n
(\vec{r}) \psi_n (\vec{r}_0)$ over the eigenfunctions $\psi_n$
satisfying $(a^2/6) \Delta \psi_n = \lambda_n \psi_n$ with the boundary
condition (\ref{eq:boundary1}).  Upon spherical integration in
(\ref{eq:flux}), all angular dependent harmonics vanish, and we arrive
at
\begin{eqnarray}
P_H(k) =  \frac{ \pi a^2 }{3 R^{\ast} r_0} \sum_{n=1}^{\infty} n
(-1)^{n+1} \ \sin \left( n \pi \frac{ r_0 }{ R^{\ast}} \right)
 \nonumber \\ \times \exp \left[ - \frac{a^2}{6} \left( \frac{n \pi}{R^{\ast}}
\right)^2 k \right] \label{eq:3din}
\end{eqnarray}

The distribution $P_P(k)$ can be derived similarly, except that now we
have to take care of the boundary condition at the outer surface of the
globule.  To this end, we argue that this condition must be taken in the
form
\begin{equation}
{\nabla}_{\vec{r}}{G( \vec{r} , k | \vec{r}_0) } |_{r=R} = 0
\label{eq:boundary2} \end{equation}
Indeed, formally this condition ensures the constant density of monomer
units throughout the globule for large values of $k$, as well as
breaking of correlations as soon as polymer chain is "reflected" by a
globular boundary.  Physically, this boundary condition reflects the
fact that there is always a "sticky layer" (or depletion layer) formed
self-consistently along the internal surface of the globule due to the
effective attraction of monomer units to the outer region where polymer
density is depleted and excluded volume effect is reduced.  As long as
we are not interested in the structure of surface layer of the globule,
we can just replace this layer by the effective boundary condition
(\ref{eq:boundary2}). After calculations for $P_P(k)$ we obtain
\begin{eqnarray}
P_P(k) = \frac{a^2 R^{\ast}}{3 (R - R^{\ast})^2 r_0} 
\sum_{n=1}^{\infty} \frac{ \zeta_n^2 \sin \left( \zeta_n \frac{r_0
- R^{\ast}}{R - R^{\ast}}\right)}{\zeta_n - \sin \zeta_n \cos
\zeta_n } \nonumber  \\ \times  \exp \left[ - \frac{a^2}{6} \left(
\frac{ \zeta_n}{R - R^{\ast}} \right)^2 k \right] \ ,
\label{eq:3dbetweengrad}
\end{eqnarray}
where $\zeta_n$ satisfies $\zeta_n = \left(1 - R^{\ast}/R \right) \tan
\zeta_n$.

Finally, to compute the dispersion $D_{\ell}$, we note that
$u_{i}u_{j}$ in eq.\ (\ref{eq:dispersion}) is the probability that
both units $i$ and $j$ are of $H$ type, which happens if both are
located inside the $R^{\ast}$ region in the parental conformation.
Thus
\begin{eqnarray}
\left< u_i u_j \right> & = & \frac{1}{V} \int_{V^{\ast}} d^3 r
\int_{V^{\ast}} d^3 r_0 G(\vec{r} , \left|i-j\right| | \vec{r}_0 )
= \nonumber \\ & = & \frac{1}{V} \sum_{n=0}^{\infty} e^{\lambda_n
\left| i - j \right|} \left( \int_{V^{\ast}} d^3 r \psi_n (\vec{r}
) \right)^2 \ , \label{eq:p_HH}
\end{eqnarray}
where $G \left( \vec{r},k|\vec{r}_0 \right)$ is the Green function
satisfying eq.\ (\ref{eq:Diffusion_equation}) with the boundary
condition (\ref{eq:boundary2}), $\psi_n$ and $\lambda_n$ are the
corresponding eigenfunctions and eigenvalues.  Plugging this into
the eq. (\ref{eq:dispersion}), one arrives at the cumbersome
looking expression for $D_{\ell}$ which is easy to implement
numerically; the result is plotted in Fig.\ 2 and shows virtually
perfect fit to the Monte Carlo data \cite{expl}. In fact, this
fit may even be somewhat fortuitous; indeed, along with the
diffusion equation (\ref{eq:Diffusion_equation}), which is an
approximation for the underlying bond fluctuation model, we can
also switch from summation to integration in eq.\
(\ref{eq:dispersion}), yielding
\begin{equation}
D_{\ell}^2 = 6 \ell^2 \sum_{n=1}^{\infty} g \left(\frac{\xi_n^2 a^2 }{6
R^2}\ell \right) \alpha_n  \ ,
\label{eq:finalD}
\end{equation}
where $\xi_n$ satisfies the equation $\xi_n = \tan \xi_n$,
$\alpha_n=\frac{\xi_n^2+1}{\xi_n^6} \left[ \frac{\xi_n
R^{\ast}}{R} \cos \frac{\xi_n R^{\ast}}{R} - \sin \frac{\xi_n
R^{\ast}}{R} \right]^2$, and the function $g$ is defined as
$g(x)=2(x-1+\exp(-x))/x^2$. (Note that the sum in eq.\
(\ref{eq:finalD}) starts from $n=1$ and does not include the
ground state, for which $\psi_0$ is a constant).  It is easy to
check that equation (\ref{eq:finalD}) does indeed have asymptotic
behavior in accord with (\ref{eq:qualitativeD}), including a broad
cross-over;  similarly, eqs.\
(\ref{eq:3din},\ref{eq:3dbetweengrad}) agree with
(\ref{eq:qualitative}).  Numerically, eq.\ (\ref{eq:finalD}) fits
pretty well to the data (Fig.\ 2), but, we repeat, best fit is
achieved by the cumbersome discrete formula.

In conclusion, we have shown that protein-like copolymers
generated according to the coloring procedure proposed earlier
\cite{PhysA,PhRev} exhibit long-range correlations in the primary
sequences.  For the flexible polymers and large enough globules,
these correlations belong to the Levy-flight statistics. This
result, first observed in computer experiments, is confirmed and
explained by analytical calculation.  Analytical theory suggests
that Levy flight statistics, albeit with a broader cross-over
region, is expected even if parental conformation is not maximally
compact, but rather a globule somewhat closer to the
$\theta$-point.  It becomes clear from our model that segregation
of globule into a hydrophobic core and a hydrophilic peel, which
is the necessary condition for water solubility, does impose
severe restrictions on the sequence, and, therefore, must be
manifested in certain correlations.  Qualitatively, this agrees
with earlier study of protein sequences \cite{ProtCor}.  However,
precise form of correlations may be affected by both globule size
and chain flexibility, including the aspect of secondary
structure. Moreover, another effect exists which favors
correlations of the opposite sign, and which dominates for small
globules and/or rigid polymers \cite{Irback}. Identification of
long range correlations in protein sequences becomes, therefore,
an interesting task promising to shed light on the evolutionary
criteria involved in the selection of proteins and the role of
water solubility among them.

The work was supported by NATO (PST/CLG. 974956), INTAS
(INTAS-OPEN-97-0678) and RFBR 98-03-33337a. The authors thank Dr.\
A.Irb\"ack for the useful correspondence.


\begin{references}
\bibitem{expl}  For the bond fluctuation model
\protect\cite{PhRev}, parameter $a$ is given by $a=\sqrt{8}b$,
where $b$ is the lattice spacing. Also, the value $R=27.8b$ was
used to adjust to the actual radius of gyration of the parent
globule at the coloring conditions.
\end{references}
\end{document}